\documentclass[twocolumn,conference]{IEEEtran}
\pdfoutput=1
\usepackage[T1]{fontenc}
\usepackage[latin9]{inputenc}
\usepackage{float}
\usepackage{url}
\usepackage{amsmath}
\usepackage{amsthm}
\usepackage{amssymb}
\usepackage{graphicx}
\usepackage[unicode=true,
 bookmarks=false,
 breaklinks=false,pdfborder={0 0 0},pdfborderstyle={},backref=false,colorlinks=false]
 {hyperref}
\hypersetup{pdftitle={Your Title},
 pdfauthor={Your Name},
 draft=True, pdfpagelayout=OneColumn, pdfnewwindow=true, pdfstartview=XYZ, plainpages=True}
\usepackage{breakurl}

\makeatletter

\floatstyle{ruled}
\newfloat{algorithm}{tbp}{loa}
\providecommand{\algorithmname}{Algorithm}
\floatname{algorithm}{\protect\algorithmname}

\theoremstyle{plain}
\newtheorem{thm}{\protect\theoremname}
\theoremstyle{plain}
\newtheorem{prop}[thm]{\protect\propositionname}
\theoremstyle{remark}
\newtheorem{claim}[thm]{\protect\claimname}

\usepackage[caption=false,font=footnotesize]{subfig}
\usepackage{algpseudocode}

\algdef{SE}{Begin}{End}{\textbf{begin}}{\textbf{end}}
\algnewcommand\algorithmicforeach{\textbf{for each}}
\algdef{S}[FOR]{ForEach}[1]{\algorithmicforeach\ #1\ \algorithmicdo}

\makeatother

\providecommand{\claimname}{Claim}
\providecommand{\propositionname}{Proposition}
\providecommand{\theoremname}{Theorem}

\begin{document}

\title{Social Welfare Maximization Auction in Edge Computing Resource Allocation
for Mobile Blockchain}

\author{Yutao Jiao, Ping Wang, Dusit Niyato, and Zehui Xiong\\
School of Computer Science and Engineering, Nanyang Technological
University, Singapore 639798}
\maketitle
\begin{abstract}
Blockchain, an emerging decentralized security system, has been applied
in many applications, such as bitcoin, smart grid, and Internet-of-Things.
However, running the mining process may cost too much energy consumption
and computing resource usage on handheld devices, which restricts
the use of blockchain in mobile environments. In this paper, we consider
deploying edge computing service to support the mobile blockchain.
We propose an auction-based edge computing resource market of the
edge computing service provider. Since there is competition among
miners, the allocative externalities (positive and negative) are taken
into account in the model. In our auction mechanism, we maximize the
social welfare while guaranteeing the truthfulness, individual rationality
and computational efficiency. Based on blockchain mining experiment
results, we define a hash power function that characterizes the probability
of successfully mining a block. Through extensive simulations, we
evaluate the performance of our auction mechanism which shows that
our edge computing resources market model can efficiently solve the
social welfare maximization problem for the edge computing service
provider.
\end{abstract}


\section{Introduction}

Decentralized cryptocurrencies have witnessed explosive growth since
the first decentralized distributed currency Bitcoin was launched
in 2009~\cite{Nakamoto2008}. In contrast to traditional currencies,
decentralized cryptocurrencies are traded among participants over
a peer-to-peer (P2P) network without relying on trusted third parties
like banks or financial regulatory authorities. As the backbone technology,
blockchain protocol provides an effective consensus mechanism to successfully
solve problems about incentive, tamper-resistance, trust and so on~\cite{Kiayias2016}.
Recently, blockchain has heralded many applications in various fields,
such as finance~\cite{Guo2016}, Internet of Things~\cite{Christidis2016},
smart grid~\cite{Kang2017} and cognitive radio~\cite{Kotobi2017}.
According to the market research firm Tractica, it is estimated that
the annual revenue for enterprise applications of blockchain will
increase to \$19.9 billion by 2025~\cite{Tractica}. It is also worth
noting that the access scope of the blockchain can be not only \emph{public}
as Bitcoin, but also \emph{private} or \emph{consortium/community}~\cite{Buterin2015}
where blockchain networks are established and usually managed by the
blockchain owner\footnote{A good example is R3 consortium (\url{https://www.r3.com/}) experimenting
an Ethereum private blockchain on \href{https://azuremarketplace.microsoft.com/en-us/marketplace/apps/microsoft-azure-blockchain.azure-blockchain-service}{Microsoft Azure Blockchain Service}.}.  

The security and reliability of blockchains depend on a distributed
consensus mechanism. Specifically, a group of participants in the
blockchain network, called \emph{miners,} try to solve a computationally
difficult problem, i.e., the \emph{proof of work} (PoW) puzzle, where
the process is called \emph{mining}. First, each miner receives and
selects certain number of transaction records from public. Once solving
the puzzle, the miner will broadcast a \emph{block} which combines
the transaction records and relevant information in the blockchain
network. Next, this block will be verified by the majority of other
miners for consensus and then finally be added to the blockchain.
The miner which successfully finishes the above steps will receive
a fixed reward and certain transaction fees as incentives of mining.

However, blockchain applications in mobile environments are still
seldom realized because solving the PoW puzzle needs high computing
power and large amount of energy which mobile devices cannot satisfy.
In this paper, we consider the edge computing services for mobile
users to deploy their mining tasks and thus support the mobile blockchain
applications. Specifically, we discuss the allocation and pricing
issue for edge computing resource. We first propose an auction-based
market model. The market consists of three entities, i.e., blockchain
owner, edge computing service provider (ESP) and miners. Considering
the competition among miners~\cite{Kiayias2016} and network effects
of blockchain by nature~\cite{Catalini2016}, we then study the auction
mechanism with allocative externalities  to maximize the social welfare.
Our social maximization mechanism is truthful (incentive compatible),
individually rational and computationally efficient. Based on our
real-world experiment of mobile blockchain, we analyze the probability
of successfully mining a block and verify the probability function.
Our simulation results show that the proposed auction market model
can not only help the ESP to make practical sale strategies, but also
incentivize the blockchain owner in adjusting the blockchain protocol.
To the best of our knowledge, this is the first work that investigates
resource management and pricing in the mobile blockchain with an auction
model.

The rest of this paper is organized as follows. Section~\ref{sec:Related-Work}
reviews related work, and the system model of edge computing resource
market for mobile blockchain is introduced in Section~\ref{sec:System-Model}.
Section~\ref{sec:Social-welfare-maximization} formulates the social
maximization problem and gives theoretical analysis. Section~\ref{sec:Experiment-and-numerical}
presents experimental results of mobile blockchain and performance
analysis. Finally, Section~\ref{sec:Conclusions} concludes the paper.

\section{Related Work\label{sec:Related-Work}}

Over the recent years, there are some studies on the economics and
mining strategies in the blockchain network. As one of the pioneer
papers, the authors in~\cite{Kroll2013} modeled the mining process
as a game played by miners. Each miner's strategy is to choose which
branch of blockchain to mine on. They proved the existence of a Nash
equilibrium when all miners behave as expected by Bitcoin designer.
Further, they explored the case that some miners may deviate from
the expected behavior, which makes blockchain network unstable and
vulnerable. The authors in~\cite{Houy2014} proposed a game model
in which the occurrence of solving the PoW puzzle is modeled as a
Poisson process. Miners have to decide the size of block to broadcast
as their response. Analytical solutions to the Nash equilibrium in
a two-miner case was given. In~\cite{Lewenberg2015}, the authors
designed a cooperative game model to investigate the mining pool.
In the pool, miners form a coalition to accumulate their computational
power and have steady reward. However, these works only studied the
internal mining scheme and paid little attention to the actual running
of blockchain in more dynamic environments, i.e., mobile blockchain.
Therefore, this motivates us to place the edge computing service as
underlying technology for mobile blockchain network and build an edge
computing resource market model. The blockchain technology can achieve
more extensive and promising applications in the mobile information
society. The auction design has been widely studied in other resource
allocation problems, such as spectrum trading~\cite{Gao2011,Zhou2008}
and data crowdsensing~\cite{Yang2016}. However, none of these works
can be directly applied to edge computing applications for mobile
blockchain, since they only focused on the specific properties constrained
by the studied topics. For example, in the blockchain networks, the
allocative externalities~\cite{Jehiel2005,Salek2008} should be considered
because of miners care about the computational power allocated to
the others. The authors in~\cite{Luong} used deep learning to recover
the classical optimal auction for revenue maximization and applied
it in the edge computing resources allocation. However, it is less
about the design of auction and cannot guarantee the incentive compatibility. 

\section{System Model: Mobile Blockchain and Market Model\label{sec:System-Model}}

\begin{figure}[tbh]
\begin{centering}
\includegraphics[width=1\columnwidth]{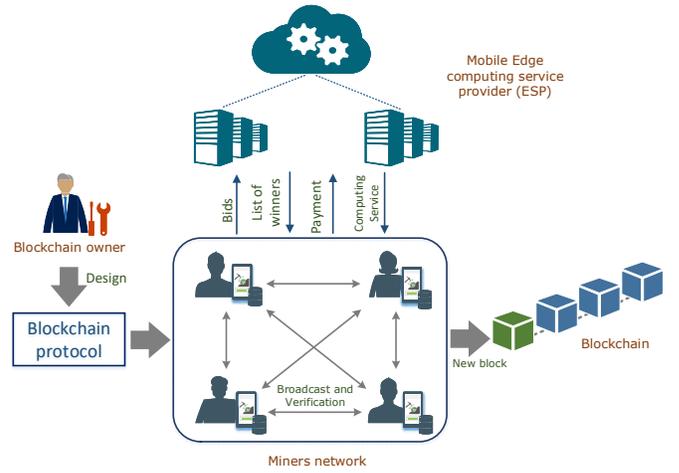}
\par\end{centering}
\caption{Edge computing resource market for mobile blockchain.\label{fig:system-model}}

\end{figure}

Figure~\ref{fig:system-model} shows the auction-based market model
for trading edge computing resources. The \emph{blockchain owner}
launches a blockchain application and designs the protocol for blockchain
network operation. The \emph{mobile users} buy resources from the
edge computing service provider (ESP) and become miners. In the miners
network, they take part in the mining process to contribute new blocks
to the blockchain.  

\subsection{Mobile Blockchain}

Blockchain can be used to develop applications with mobile devices,
as indicated in our earlier study~\cite{Suankaewmanee2018}. To support
the blockchain based service, there are a set of miners continuously
running a consensus protocol~\cite{Nakamoto2008} to confirm and
secure distributed data or transactions at backend. According to the
protocol, miners are required to finish the mining task, i.e., solving
the PoW puzzle. The mining process is conducted in a tournament structure,
and miners chase each other to obtain the solution. Specifically,
the PoW algorithm involves finding a nonce value that, together with
additional fields about all valid and received transactions, the previous
block and the timestamp, the output satisfies a given condition. If
the nonce is found, the miner will combine it and additional fields
into a block and then broadcast the block to peers in the blockchain
network for verification and reaching consensus. Finally, the new
block can be linked to the existing accepted chain of blocks. However,
for a mobile user, it is unrealistic to continuously run such a computationally
difficult program which requires high computing power and consumes
a large volume of energy and time. Because the outstanding characteristics
of edge computing: low latency, mobility and wide-spread geographical
distribution~\cite{Ahmed2016}, we consider offloading the mining
tasks to the edge servers. 

\subsection{Edge Computing Resources Trading \label{subsec:Edge-Computing-Resources}}

As shown in Fig.~\ref{fig:system-model}, we consider a scenario
where there is one ESP, one blockchain owner and a community of mobile
users $\mathcal{N=}\{1,\ldots,N\}$. Each mobile user wants to be
a miner which runs a mobile blockchain application to record and verify
the transactions or data sent to the community. Due to the computing
limitation on their devices, mobile users want to offload the task
of solving PoW to the nearby edge computing servers deployed by the
ESP. In particular, the ESP launches an auction to sell its services.
It first announces its service and relevant information to mobile
users. Then, the mobile users submit their resource demand profile
$\mathbf{d}=(d_{1},\ldots,d_{N})$ and corresponding bids $\mathbf{b}=(b_{1},\ldots,b_{N})$
which represents their valuations of the offered services. After receiving
the demands and bids, the ESP selects the winners as the successful
miners and notifies the mobile users the allocation $\mathbf{x}=(x_{1},\ldots,x_{N})$
and the service price $\mathbf{p}=(p_{1},\ldots,p_{N})$. The setting
$x_{i}=1$ means user $i$ is within the winner list and being allocated
resources that it demands for while $x_{i}=0$ is for no resource\footnote{The user becomes a miners if it wins the auction.}.
$p_{i}$ is the sale price that user $i$ is charged by the ESP\footnote{The payment for user which is not allocated any resource is zero,
i.e., $p=0$.}. At the end of the auction, the winners make the payment and access
the edge computing service. 

\subsection{Blockchain Mining with Edge Computing Service \label{subsec:Blockchain-Mining-with-ES} }

With the allocation $x_{i}$ and demand $d_{i}$, miner $i$'s hash
power $\gamma_{i}$ relative to other miners' allocated resources
can be calculated by: 

\begin{equation}
\gamma_{i}(\mathbf{d},\mathbf{x})=\frac{d_{i}^{\alpha}x_{i}}{\sum_{j\in\mathcal{N}}d_{j}^{\alpha}x_{j}}\label{eq:hash-power}
\end{equation}
which is a fraction function that $\sum_{i\in\mathcal{N}}\gamma_{i}=1.$
$\alpha$ is the curve fitting parameter of the hash power function
$\gamma_{i}(\mathbf{d},\mathbf{x})$ verified by our real-world experiment,
the detail of which will be presented in Section~\ref{sec:Experiment-and-numerical}. 

In the mining tournament, miners compete to be the first to solve
PoW with correct nonce value and propagate the block to reach consensus.
The generation of new blocks follows a Poisson process with a constant
rate $\frac{1}{\lambda}$ throughout the whole blockchain network~\cite{Kraft2016}.
Before the tournament, miners collect unconfirmed transactions into
their blocks. We represent the size of transactions of each miner
by $\mathbf{s}=(s_{1},\ldots,s_{N})$. When miner $i$ propagates
its block to the mobile blockchain network for consensus, the time
for verifying each transaction is affected by the size of transactions
$s_{i}$. The first miner which successfully has its block achieve
consensus can get a reward $R$. The reward is composed of a fixed
bonus $T$ for mining a new block and a flexible transaction fee $t$
determined by the size of its collected transactions $s$ and the
transaction fee rate $r$~\cite{Houy2014}. Thus, miner $i$'s expected
reward $R_{i}$ can be expressed by:

\begin{equation}
R_{i}=(T+rs_{i})\mathbb{P}_{i}(\gamma_{i}(\mathbf{d},\mathbf{x}),s_{i}),\label{eq:Expected-reward}
\end{equation}
where $\mathbb{P}_{i}(\gamma_{i}(\mathbf{d},\mathbf{x}),s_{i})$ is
the probability that miner $i$ receives the reward by contributing
a block to the blockchain. 

From the mining tournament above, winning the reward depends on the
successful mining and instant propagation. The probability of mining
a new block $P_{i}^{m}$ is equal to miner $i$'s hash power $\gamma_{i}$,
i.e., $P_{i}^{m}=\gamma_{i}$. However, the miner may even lose the
tournament if its new block does not achieve consensus as the first.
This kind of mined block that cannot be added on to the blockchain
is called orphaned block~\cite{Houy2014}. Moreover, the block containing
larger size of transactions has higher chance becoming orphaned. This
is because a larger block needs more propagation time, thus causing
higher delay for consensus. Here, we assume miner $i$'s block propagation
time $\tau_{i}$ is linear to the size of transactions in its block,
i.e., $\tau_{i}=\xi s_{i}$. $\xi$ is a constant that reflects the
impact of $s_{i}$ on $\tau_{i}$. Since the arrival of new blocks
follows a Poisson distribution, miner $i$'s orphaning probability
can be approximated as follows~\cite{Rizun2015}:

\begin{equation}
P_{i}^{o}=1-\exp(-\frac{1}{\lambda}\tau_{i}).\label{eq:p-orphan}
\end{equation}
After substituting $\tau_{i}$, we can express $\mathbb{P}_{i}$ as
follows:

\begin{align}
\mathbb{P}_{i} & =P_{i}^{m}(1-P_{i}^{o})\label{eq:p-mining-a-block}\\
 & =\gamma_{i}e^{-\frac{1}{\lambda}\xi s_{i}}.\nonumber 
\end{align}

\subsection{Blockchain Management}

The blockchain owner maintains the blockchain mining protocol that
specifies the fixed bonus $T$ for the contributing miner and the
transaction fee rate $r$. Through adjusting the difficulty of finding
a new block, the blockchain owner keeps the average time $\lambda$
at a reasonable constant value\footnote{It is worth noting that the blockchain owner is not a central entity
that controls the data storage and mining strategies of the miners.
Similar to the bitcoin protocol~\cite{Nakamoto2008}, the blockchain
owner in this paper only designs the protocol specifies the value
of $T$, $r$ and $\lambda$, and does not affect the decentralization
and security of blockchain. }. Additionally, a blockchain in PoW systems is only as secure as the
amount of computing power dedicated to mining it~\cite{Catalini2016}.
This results in positive network effects: as more mobile users participate
in mining and more computing resources are invested, the value of
reward given to miners increases since the blockchain network is more
stable and secure. Empirically, we define the network effects by a
common S-shaped utility function~\cite{Jackson2010}:

\begin{equation}
w(d_{\mathcal{N}})=\frac{1-e^{-\nu d_{\mathcal{N}}}}{1+\mu e^{-\nu d_{\mathcal{N}}}},\label{eq:S-shape}
\end{equation}
where $d_{\mathcal{N}}=\sum_{i\in\mathcal{N}}d_{i}x_{i}$ is the total
quantity of allocated resources and $\mu,\nu$ are positive parameters.
The monotonic increase of network effect function begins slowly from
$0$, then accelerates (convexly), and then eventually slows down
(concavely) and converges asymptotically to $1$.

\section{Social Welfare Maximization Auction for Edge Computing Service\label{sec:Social-welfare-maximization}}

In this section, we propose an auction mechanism for the ESP to allocate
edge computing resources efficiently. We focus on maximizing the social
welfare while guaranteeing the truthfulness, individual rationality
and computational efficiency.

\subsection{Valuation of mobile users}

To take part in the auction, a mobile user needs to give the bid representing
its valuation to the auctioneer, i.e., the ESP. Since the mobile user
$i$ cannot know the number of winners and total supply of computing
resources until auction ends, it can only give the bid $b_{i}$ according
to its expected reward $R_{i}$ which is also called \emph{ex-ante}
valuation $v_{i}^{'}$, i.e., 

\begin{equation}
v_{i}^{'}=R_{i}\label{eq:ex-ante-valuation-o}
\end{equation}

After the auction result is released, user $i$ has an \emph{ex-post}
valuation $v_{i}^{''}$ of the edge computing service considering
network effects, which is defined by 

\begin{equation}
v_{i}^{''}=R_{i}w\label{eq:ex-post-valuation-o}
\end{equation}
 where $w$ is the network effect defined in~(\ref{eq:S-shape}).

After substituting (\ref{eq:hash-power}), (\ref{eq:p-mining-a-block}),
(\ref{eq:S-shape}) into (\ref{eq:ex-ante-valuation-o}) and (\ref{eq:ex-post-valuation-o}),
we have the specific expression of user $i$'s ex-ante and ex-post
valuation:

\begin{equation}
v_{i}^{'}=(T+rs_{i})e^{-\frac{1}{\lambda}\xi s_{i}},\label{eq:ex-ante}
\end{equation}

\begin{equation}
v_{i}^{''}=\frac{d_{i}^{\alpha}x_{i}}{\sum_{j\in\mathcal{N}}d_{j}^{\alpha}x_{j}}\frac{1-e^{-\nu\sum_{i\in\mathcal{N}}d_{i}x_{i}}}{1+\mu e^{-\nu\sum_{i\in\mathcal{N}}d_{i}x_{i}}}(T+rs_{i})e^{-\frac{1}{\lambda}\xi s_{i}}.\label{eq:ex-post}
\end{equation}

\subsection{Auction Maximizing Social Welfare}

Once receiving bids $\mathbf{b}$ from all the mobile users, ESP will
select winners and determine corresponding payments to maximize the
social welfare. Let $c$ denote the unit cost of running the edge
computing service. ESP's total cost is $C(d_{\mathcal{N}})=cd_{\mathcal{N}}$.
Thus, designing such an auction becomes solving an optimization problem:

\begin{align}
\max_{\mathbf{x}} & \sum_{i\in\mathcal{N}}\frac{d_{i}^{\alpha}x_{i}}{\sum_{j\in\mathcal{N}}d_{j}^{\alpha}x_{j}}\frac{1-e^{-\nu\sum_{i\in\mathcal{N}}d_{i}x_{i}}}{1+\mu e^{-\nu\sum_{i\in\mathcal{N}}d_{i}x_{i}}}(T+rs_{i})e^{-\frac{1}{\lambda}\xi s_{i}}\nonumber \\
 & \quad-\sum_{i\in\mathcal{N}}cd_{i}x_{i}\label{eq:socail-welfare-original}\\
s.t. & \sum_{i\in\mathcal{N}}d_{i}x_{i}\leq D\label{eq:constraint-supply}\\
 & x_{i}\in\{0,1\},\forall i\in\mathcal{N}
\end{align}
where the objective function in (\ref{eq:socail-welfare-original})
is the difference between the sum of all users' ex-post valuations
and ESP's total cost. The constraint in (\ref{eq:constraint-supply})
defines the maximum quantity of computing resources that ESP can offer
denoted by $D$. 

Based on the above system model, we first consider a simpler case
where all mobile users submit various bids to compete for a fixed
quantity of resources. Without loss of generality, we set $d_{i}=1,\forall i\in\mathcal{N}$.
Then, the optimization problem can be expressed as follows:

\begin{align}
\max_{\mathbf{x}} & \sum_{i\in\mathcal{N}}\frac{x_{i}}{\sum_{j\in\mathcal{N}}x_{j}}\frac{1-e^{-\nu\sum_{i\in\mathcal{N}}x_{i}}}{1+\mu e^{-\nu\sum_{i\in\mathcal{N}}x_{i}}}(T+rs_{i})e^{-\frac{1}{\lambda}\xi s_{i}}\nonumber \\
 & \quad-\sum_{i\in\mathcal{N}}cx_{i}\label{eq:socail-welfare-original-unit}\\
s.t. & \sum_{i\in\mathcal{N}}x_{i}\leq D\label{eq:constraint-supply-1}\\
 & x_{i}\in\{0,1\},\forall i\in\mathcal{N}\label{eq:xi_integer}
\end{align}

We aim to solve this integer programming efficiently while making
the auction process truthful and individually rational. The proposed
auction is based on the Myerson's well-known characterization~\cite{Myerson1981}
as described in Theorem~\ref{thm: truthful-condition}. 
\begin{thm}
(\cite[Theorem~13.6]{Nisan2007}) An auction is truthful if and only
if it satisfies the following two properties:\label{thm: truthful-condition}
\end{thm}
\begin{enumerate}
\item \emph{Monotonicity of winner selection rule: If user $i$ wins the
auction with bid $b_{i}$, then it will also win with any higher bid
$b_{i}'>b_{i}$.}
\item \emph{Critical payment: The payment by a winner is the smallest value
needed in order to win the auction. }
\end{enumerate}
By using Theorem~\ref{thm: truthful-condition}, our auction mechanism
is illustrated in Algorithm~\ref{alg:1}. In Lines~5-12, the winner
selection process is conducted with a greedy scheme. We define a winner
set $\mathcal{W}$. Including a user $i$ in the set is equivalent
to assigning $x_{i}=1$. Thus, we rewrite the problem in an alternative
form as follows:

\begin{align*}
\max_{\mathcal{W\subseteq N}} & S(\mathcal{W})\\
s.t. & S(\mathcal{W})=\sum_{i\in\mathcal{W}}\frac{1}{\mathcal{\left|W\right|}}\frac{1-e^{-\nu\left|\mathcal{W}\right|}}{1+\mu e^{-\nu\left|\mathcal{W}\right|}}b_{i}-c\left|\mathcal{W}\right|,\\
 & \mathcal{\left|W\right|}\leq D
\end{align*}
where $\left|\mathcal{W}\right|$ measures the number of winners in
$\mathcal{W}$, $S(\mathcal{W})$ is social welfare of $\mathcal{W}$
and $b_{i}=v_{i}'=(T+rs_{i})e^{-\lambda ks_{i}}$ because the auction
is truthful. In the winner selection process (Lines~6-12), mobile
users are first sorted in a descending order according to their bids.
We then add one user sequentially to the winner set $\mathcal{W}$,
which will be stopped before the corresponding social welfare $S(\mathcal{W})$
decreases. Finally, the solution $\mathcal{W}$ is output by the algorithm. 
\begin{prop}
The resource allocation $\mathbf{x}$ output by Algorithm~\ref{alg:1}
is globally optimal to the social welfare maximization problem given
in (\ref{eq:socail-welfare-original-unit})-(\ref{eq:xi_integer}).\label{prop:resource-allocation}
\end{prop}
\begin{IEEEproof}
With proof by contradiction, this result follows from Claim~\ref{claim:3}.
\end{IEEEproof}
\begin{claim}
Let $\mathcal{W}_{A}$ be the solution output by Algorithm~\ref{alg:1}
on input $\mathbf{b}$, and $\mathcal{W}_{O}$ the optimal solution.
If $\mathcal{W}_{A}\neq\mathcal{W}_{O}$, then we can construct another
solution $\mathcal{W}_{O}^{*}$ the social welfare of which $S(\mathcal{W}_{O}^{*})$
is even larger than $\mathcal{W}_{O}$.\label{claim:3}
\end{claim}
\begin{IEEEproof}
Without loss of generality, we assume $b_{1}\geq\cdots\geq b_{N}$
and $\mathcal{W}_{A}\neq\mathcal{W}_{O}$. Let $m$ be the first
element (while-loop Lines~6-12) where $m\notin\mathcal{W}_{O}$.
Since ($b_{m}$ is minimal by assumption.) $m$ is maximal, we must
have $1,\ldots,m-1\in\mathcal{W}_{O}$ and in particular, the set
of corresponding bid $\mathbf{b}_{\mathcal{W}_{O}}$ has the form
$\mathbf{b}_{\mathcal{W}_{O}}=\{b_{1},b_{2},\ldots,b_{m-1},b'_{m},b'_{m+1},\ldots,b'_{\left|\mathcal{W}_{O}\right|}\}$,
where the bids $b_{1},\ldots,b'_{\left|\mathcal{W}_{O}\right|}$ are
listed in the descending order. Meanwhile, Algorithm~\ref{alg:1}
chooses $\mathbf{b}_{\mathcal{W}_{A}}=\{b_{1},b_{2},\ldots,b_{m-1},b{}_{m},b{}_{m+1},\ldots,b{}_{\left|\mathcal{W}_{O}\right|}\}$
and there must be $b_{m}>b'_{j}$ for all $j\geq m$. In particular,
we have $b_{m}>b'_{m}$. Hence, we define $\mathbf{b}_{\mathcal{W}_{O}^{*}}=\mathbf{b}_{\mathcal{W}_{O}}\cup\{b_{m}\}\setminus\{b'_{m}\}$
, i.e., we obtain $\mathbf{b}_{\mathcal{W}_{O}^{*}}$ by deleting
the $m$th bid in $\mathbf{b}_{\mathcal{W}_{O}}$ and adding $b_{m}$.
Now we have the social welfare of $\mathbf{b}_{\mathcal{W}_{O}^{*}}$:

\[
S(\mathcal{W}_{O}^{*})=S(\mathcal{W}_{O})+\frac{1}{\left|\mathcal{W}_{O}\right|}\frac{1-e^{-\nu\left|\mathcal{W}_{O}\right|}}{1+\mu e^{-\nu\left|\mathcal{W}_{O}\right|}}(b_{m}-b'_{m}).
\]
Since $b_{m}-b'_{m}>0$ and $\left|\mathcal{W}_{O}^{*}\right|=\left|\mathcal{W}_{O}\right|$,
$S(\mathcal{W}_{O}^{*})$ is strictly larger than $S(\mathcal{W}_{O})$,
which is in contradiction to that $\mathcal{W}_{O}$ is the optimal
solution. This proves the claim.
\end{IEEEproof}
In Lines~13-24, for each iteration, we exclude one winner from the
user set and rerun the winner selection process to calculate the payment
for the winner. The payment calculation is based on the well-known
Vickrey\textendash Clarke\textendash Groves (VCG) mechanism~\cite{Krishna2009}. 

\begin{algorithm}[tbh]
\begin{algorithmic}[1]
\Require{Mobile users' bid profile~$\mathbf{b}=({b_{1}},\ldots,{b_{N}})$.}
\Ensure{Resource allocation profile $\mathbf{x}=({x_{1}},\ldots,{x_{N}})$ and payment profile~$\mathbf{p}=({p_{1}},\ldots,{p_{N}})$.}
\Begin
	\ForEach{$i \in \mathcal{N}$}
		\State{$x_i \gets 0, p_i \gets 0$}
	\EndFor
	\State{$\mathcal{W}\gets \varnothing, \mathcal{W}_{t}\gets \varnothing, S \gets 0, S_t \gets 0$}
	
	\Comment{$\mathcal{W}$ is the set of winners.}
	\While{$S \leq S', \mathcal{W} \neq \mathcal{N}, {\left|\mathcal{W}\right|} \leq D$}
		\State{$j \gets {\arg\max}_{j \in \mathcal{N}\setminus\mathcal{W}}b_{j}$}
		\State{$\mathcal{W} \gets \mathcal{W}_{t}$}
		\State{$\mathcal{W}_{t} \gets \mathcal{W} \cup \{j\}, S \gets S_t$}
\State{$S_t\gets\frac{1}{\left|\mathcal{W}_{t}\right|}\frac{1-e^{-\nu\left|\mathcal{W}_{t}\right|}}{1+\mu e^{-\nu \left|\mathcal{W}_{t}\right|}}\sum_{l\in\mathcal{W}_{t}}b_l-c\left|\mathcal{W}_{t}\right|$}

	\EndWhile
	\ForEach{$j \in \mathcal{W}$}
		\State{$x_j \gets 1$}
		\State{$\mathcal{N}_{-j} \gets \mathcal{N} \setminus \{j\}, \mathcal{W}_{-j} \gets \mathcal{W} \setminus \{j\}$}
	\State{$\mathcal{W}' \gets \varnothing, \mathcal{W}_{t}' \gets \varnothing, S' \gets 0, S_{t}' \gets 0$}
	\While{$S' \leq S_{t}', \mathcal{W}' \neq \mathcal{N}_{-j}$}
		\State{$k \gets {\arg\max}_{k \in \mathcal{N}_{-j}\setminus\mathcal{W}'}b_k$}
		\State{$\mathcal{W}' \gets \mathcal{W}_{t}'$}
		\State{$\mathcal{W'}_{t} \gets \mathcal{W}' \cup \{k\}, S' \gets S_{t}'$}
\State{$S_{t}'\gets\frac{1}{\left|\mathcal{W}_{t}'\right|}\frac{1-e^{-\nu\left|\mathcal{W}_{t}'\right|}}{1+\mu e^{-\nu \left|\mathcal{W}_{t}'\right|}}\sum_{l\in\mathcal{W}_{t}'}b_l-c\left|\mathcal{W}_{t}'\right|$}
	\EndWhile
\State{$p_j=S'-\frac{1}{\left|\mathcal{W}_{-j}\right|}\frac{1-e^{-\nu\left|\mathcal{W}_{-j}\right|}}{1+\mu e^{-\nu \left|\mathcal{W}_{-j}\right|}}\sum_{l\in\mathcal{W}_{-j}}b_l+c\left|\mathcal{W}_{-j}\right|$}
	\EndFor

\End
\end{algorithmic} 

\caption{Social Welfare Maximization Auction \label{alg:1}}

\end{algorithm}
\begin{figure*}[t]
\begin{centering}
\par\end{centering}
\begin{equation}
S(\{b_{1},\ldots b_{k-1},b_{k}\})=\frac{1-\mathrm{e}^{-\nu k}}{(1+\mu\mathrm{e}^{-\nu k})k}\left(\sum_{j=1}^{k-1}b_{j}+b_{k}\right)<\frac{1-\mathrm{e}^{-\nu k}}{(1+\mu\mathrm{e}^{-\nu k})k}\left(\sum_{j=1}^{k-1}b_{j}+b_{i}^{+}\right)=S\left(\{b_{1},\ldots,b_{k-1},b_{i}^{+}\}\right)\label{eq:inequality}
\end{equation}
\end{figure*}
\begin{prop}
The Social Welfare Maximization Auction (Algorithm~\ref{alg:1})
is truthful.
\end{prop}
\begin{IEEEproof}
Since the calculation of payment by the algorithm relies on VCG mechanism,
it directly satisfies the second condition in Theorem~\ref{thm: truthful-condition}.
For the first condition about monotonicity, we only need to show that
if a winner $i$ raises its bid from $b_{i}$ to $b_{i}^{+}$where
$b_{i}^{+}>b_{i}$, it still stays in the set of winners. We denote
the original set of winners as $\mathcal{W}$ and the new set of winners
$\mathcal{W}_{+}$ after winner $i$ changes its bid to $b_{i}^{+}$.
The original bid set is $\mathbf{b}=\{b_{1},\ldots,b_{i},\ldots b_{N}\}$
$(i\leq\mathcal{\left|W\right|})$ sorted in the descending order.
In addition, we define $S(\mathbf{b}_{\mathcal{U}})=S(\mathcal{U}),\forall\mathcal{U}\subseteq\mathcal{N}$
which means the social welfare of a set of bids is equal to the set
of its corresponding users. We discuss the monotonicity in two cases:

1) Case 1: $b_{i-1}\geq b_{i}^{+}\geq b_{i}\geq b_{i+1}$. The new
set of ordered bids is $\mathbf{b}^{+}=\{b_{1},\ldots,b_{i-1},b_{i}^{+},b_{i+1},\ldots b_{N}\}$.
We have 
\begin{align}
S(\{b_{1},\ldots,b_{i}^{+}\})= & \frac{1-e^{-\nu i}}{(1+\mu e^{-\nu i})i}\left(\sum_{j=1}^{i-1}b_{j}+b_{i}^{+}\right)-ci\nonumber \\
> & S(\{b_{1},\ldots,b_{i}\})=\sum_{j=1}^{i}\frac{1-e^{-\nu i}}{\left(1+\mu e^{-\nu i}\right)i}b_{j}-ci.\label{eq:S-case1}
\end{align}
The social welfare of new set of bids $\{b_{1},\ldots,b_{i}^{+}\}$
is larger than that of original set of bids $\{b_{1},\ldots,b_{i}\}$,
which guarantees $b_{i}^{+}$ being in the set of winning bids.

2) Case 2: $b_{k-1}\geq b_{i}^{+}\geq b_{k}\geq\ldots\geq b_{i}$,
$1<k<i$. The new set of ordered bids is $\mathbf{b}^{+}=\{b_{1},\ldots,b_{k-1},b_{i}^{+},b_{k},\ldots,b_{i+1},\ldots b_{N}\}$.
We have 
\begin{equation}
S(\{b_{1},\ldots,b_{k-1},b_{i}^{+}\})=\frac{1-\mathrm{e}^{-\nu k}}{(1+\mu\mathrm{e}^{-\nu k})k}\left(\sum_{j=1}^{k-1}b_{j}+b_{i}^{+}\right)-ck,\label{eq:S-case2-k}
\end{equation}

\begin{equation}
S(\{b_{1},\ldots b_{k-1},b_{k}\})=\frac{1-\mathrm{e}^{-\nu k}}{(1+\mu\mathrm{e}^{-\nu k})k}\sum_{j=1}^{k}b_{j}-ck,\label{eq:S-case-original-k}
\end{equation}

\begin{equation}
S(\{b_{1},\ldots,b_{k-1}\})=\frac{1-\mathrm{e}^{-\nu(k-1)}}{(1+\mu\mathrm{e}^{-\nu(k-1)})(k-1)}\sum_{j=1}^{k-1}b_{j}-c(k-1).\label{eq:S-case2-k-1}
\end{equation}

Since in the original set of bids $\mathbf{b}$, $\{b_{1,\ldots,}b_{k-1},b_{k},\ldots,b_{i}\}$
are all selected as winning bids, $S(\{b_{1},\ldots b_{k-1},b_{k}\})>S(\{b_{1},\ldots,b_{k-1}\})$.
Because the inequality in (\ref{eq:inequality}), we have 

\[
S(\{b_{1},\ldots,b_{k-1},b_{i}^{+}\})>S(\{b_{1},\ldots,b_{k-1}\}),
\]
which implies that $b_{i}^{+}$still wins the auction. This concludes
the proof.
\end{IEEEproof}
\begin{prop}
The Social Welfare Maximization Auction (Algorithm~\ref{alg:1})
is computationally efficient and individually rational.
\end{prop}
\begin{IEEEproof}
Since the time complexity of finding the maximum miner's bid is $O(N)$
and the number of winners is at most $N$, the time complexity of
the winner selection process (while-loop, Lines~6-12) is $O(N^{2})$.
In each iteration of payment calculation process (Lines~13-24), a
similar winner selection process is executed. Therefore, the whole
auction process can be performed in polynomial time with the time
complexity of $O(N^{3})$ which is efficient. According to Proposition~\ref{prop:resource-allocation}
and the properties of the VCG mechanism~\cite{Krishna2009}, the
payment scheme in Algorithm~\ref{alg:1} guarantees the individual
rationality. 
\end{IEEEproof}

\section{Experiment results and performance analysis\label{sec:Experiment-and-numerical}}

In this section, we provide simulation results of the proposed auction,
from which we can further obtain useful decision making strategies
for ESP and the blockchain owner. 

\subsection{Verification for Hash Power Function}

\begin{figure}[tbh]
\begin{centering}
\includegraphics[width=0.65\columnwidth]{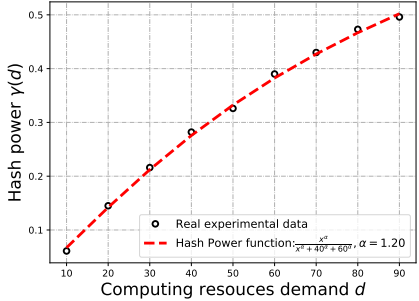}
\par\end{centering}
\caption{Estimation of the hash power function $\gamma(d)$\label{fig::hash-power}.}

\end{figure}

An earlier real-world mobile blockchain mining experiment has been
done in~\cite{Suankaewmanee2018,Xiong2017}. In the experiment, we
designed a mobile blockchain client application in the Android platform
and implemented it on three mobile devices (miners). Each of the three
client applications generates transactions and then starts mining
with one CPU core. The miners' CPU utilization rate is managed and
measured on the Docker platform~\cite{docker}. Each mobile device
mines the block under Go-Ethereum~\cite{Go-ethereum} blockchain
framework. To verify the hash power function (\ref{eq:hash-power}),
we vary one miner\textquoteright s service demand while fixing the
other two miners\textquoteright{} service demand (CPU utilization)
at 40 and 60. Besides, we set the number of transactions in each mined
block to be 10 for all miners. Figure~\ref{fig::hash-power} shows
the change of the hash power, i.e., the probability of successfully
mining a block with different amount of computing resources. We note
that the hash power function defined in (\ref{eq:hash-power}) can
well fit the actual experimental data. From these results, we choose
the hash power function with parameter $\alpha=1.2$ in the rest of
this section.

\subsection{Simulation Results}

We vary the number of mobiles users $N$ from $100$ to $1000$, the
mining bonus $T$ from $0$ to $5$, and the transaction fee rate
$r$ from $0.001$ to $0.009$. We set $\mu=0.5$, $\nu=0.005$, $\xi=1$
and $c=0.02$. The transaction size $s$ of each user is uniformly
distributed over $[0,1000]$. Since the blockchain owner can adjust
the average time of mining a block, we also varied the average time
of mining a block $\lambda$ from $100$ to $1800$ with increment
of $212.5$. Each measurement is averaged over $100$ instances.

\begin{figure}[tbh]
\begin{centering}
\includegraphics[width=0.65\columnwidth]{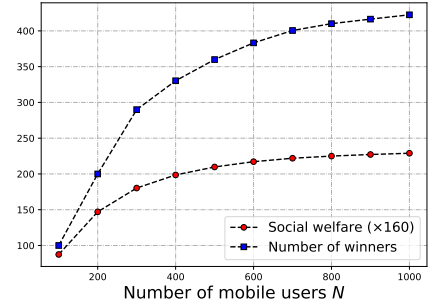}
\par\end{centering}
\caption{Impact of the number of mobile users on social welfare $S$ and number
of winners $\left|\mathcal{W}\right|$.\label{fig:N}}

\end{figure}

\begin{figure}[tbh]
\begin{centering}
\includegraphics[width=0.65\columnwidth]{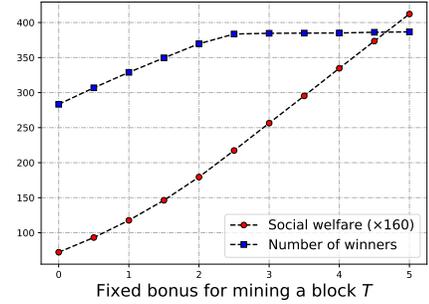}
\par\end{centering}
\caption{Impact of fixed bonus $T$.\label{fig:T}}

\end{figure}

\begin{figure}[tbh]
\begin{centering}
\includegraphics[width=0.65\columnwidth]{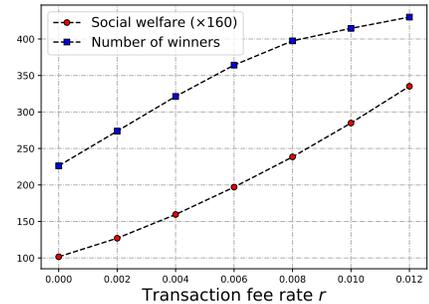}
\par\end{centering}
\caption{Impact of the transaction fee rate $r$.\label{fig:r}}

\end{figure}

\begin{figure}[tbh]
\begin{centering}
\includegraphics[width=0.65\columnwidth]{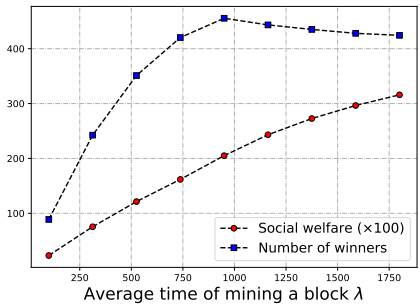}
\par\end{centering}
\caption{Impact of the parameter $\lambda$.\label{fig:lambda}}

\end{figure}
\begin{enumerate}
\item \emph{Impact of the number of mobile users $N$:} Figure~\ref{fig:N}
shows the impact of the total number of mobile users $N$ on the social
welfare $S$ and the number of participating users $\left|\mathcal{W}\right|$.
We fix $T=2.5$, $r=0.007$ and $\lambda=600$. We observe that $\left|\mathcal{W}\right|$
and $S$ increase at diminishing rate as the base of mobile users
becomes larger. Naturally, the ESP can select more winners as miners
to increase the social welfare with more mobiles users. However, at
the same time, the negative effects from the competition among a larger
number of miners are apparent, which slows down the rise of the social
welfare as well as the number of winners.
\item Impact of the fixed bonus $T$ for mining a block and the transaction
fee rate $r$: We set $N=600$ and $\lambda=600$. By fixing $r=0.007$
and $T=2.5$, we consider the impact of varied fixed bonus and transaction
fee rate on the social welfare and the number of selected miners.
From Figs.~\ref{fig:T} and~\ref{fig:r}, we note that as if the
blockchain owner raises the bonus or the transaction fee rate, more
social welfare will be generated in nearly proportion. However, the
number of winners increases and tends to be stable. This is because
there will be fierce competition if too many miners participate in
the blockchain network, which causes the loss of social welfare. 
\item Impact of the average time $\lambda$ for successfully mining a block:
In Fig.~\ref{fig:lambda}, we fix $N=600$, $T=2.5$ and $r=0.007$.
When the blockchain owner raises the difficulty of mining a block,
represented by $\lambda$, the social welfare increases while the
number of winners initially increases and then declines. Note that
the user's expected reward $R$, i.e., the valuation for edge computing
service, grows with increasing $\lambda$. When the difficulty $\lambda$
is small and each user's valuation is also small, the ESP has to accept
more users, i.e., more winners, to maximize the social welfare. However,
if the difficulty of mining a block becomes high and each user values
the service more, the ESP can reduce the number of winning users while
achieving the optimal social welfare. Another reason for the decreasing
number of winners is the increasingly intense competition among them. 
\end{enumerate}

\section{Conclusions\label{sec:Conclusions}}

In this paper, we have investigated the edge computing services that
enable mobile blockchain. To efficiently allocate computing resources,
we have proposed an auction-based market model to maximize the social
welfare. In the auction design, we have considered allocative externalities,
including the competition among the miners as well as the network
effects in the blockchain network. By theoretical analysis and simulation,
we have proved that the auction mechanism is truthful, individually
rational and computationally efficient and solves the social welfare
maximization problem. For the future work, we will consider variable
demands and corresponding bids of mobile users.

\bibliographystyle{ieeetr}
\bibliography{IEEEabrv,PC3_ESP_Blockchain_v3}

\end{document}